# Bio-inspired Synthetic Ivory as a Sustainable Material for Piano Keys

**Dieter Fischer \*, Sarah C. Parks and Jochen Mannhart**

Max Planck Institute for Solid State Research, Heisenbergstrasse 1, 70569 Stuttgart, Germany
\* Correspondence: d.fischer@fkf.mpg.de

**Abstract:** Natural ivory is no longer readily or legally available, as it is obtained primarily from elephant tusks, which now enjoy international protection. Ivory, however, is the best material known for piano keys. We present a hydroxylapatite-gelatin biocomposite that is chemically identical to natural ivory but with functional properties optimized to replace it. As this biocomposite is fabricated from abundant materials in an environmentally friendly process and is furthermore biodegradable, it is a sustainable solution for piano keys with the ideal functional properties of natural ivory.

## 1. Introduction

For centuries, ivory has been the material of choice for the veneer of piano keys. The continuing preference for ivory is driven by its excellent tactile properties. Ivory features a rather high thermal conductivity and provides a desirable grip for the fingers. These tactile properties even endure extensive playing due to ivory's significant moisture absorption capacity. Furthermore, the optical appearance of polished ivory is appealing because of its characteristic soft, warm color and luster. In addition to piano keys, ivory has been used for art objects, everyday items such as handles and billiard balls, and for the mouthpieces of certain wind instruments.

Ivory is obtained primarily from the tusks of African elephants. The global demand for ivory has caused the population of African forest elephants to shrink in the past decade by 60%, mainly due to poaching [1, 2]. The urgency to protect these animals [3] is reflected in the international trade ban for ivory. However, as long as there is a demand, elephants will continue to be killed for their tusks. This has motivated us to develop an appropriate alternative material.

Further sources of ivory are the tusks or teeth of walruses, whales and warthogs [4, 5]. Numerous alternative materials, some of which have been dubbed "artificial ivory", have been used over the years, including bones, shells, the beak or "casque" of the endangered helmeted hornbill and even the nuts of Phytelephas palm trees, called Tagua nuts or Jarine seeds. Manufactured materials used as ivory substitutes range from celluloid to mixtures of resin and casein or ivory dust to polyester or phenolic resins, see for example [4, 6-10]. The ivory substitutes used for piano keys include plastics, mineral-resin composites and bones. However, none of these materials feature all the functional properties of natural ivory.

We therefore set out to fabricate a chemically equivalent artificial material having identical—or possibly even improved—functional properties as natural ivory for the use case of piano keys. The chemical equivalence to natural ivory ensures that the material is environmentally friendly and biodegradable. For sustainability reasons, the new material has to be composed of abundant resources to be economically viable. At the same time, it has to be clearly identifiable by common analytical techniques in order not to fall under the trade ban on natural ivory.

Natural ivory is a biocomposite material [11] with a complex hierarchical architecture. Its main component is dentin, a mineralized tissue consisting of mineral material embedded in an organic matrix of collagenous proteins and water [12, 13]. The mineral portion is carbonate-enriched hydroxylapatite, which is characterized by the general formula $Ca_{10}[(PO_4)_6(CO_3)]H_2O$ with possible admixtures of further ions such as $Mg^{2+}$ [14, 15]. These rod-shaped apatite crystals (~20–100 nm in length) surround the fibril-forming collagen molecules (~300 nm in length), which are oriented along the length of the tusk (see Figure 1). The microscopic structure of ivory includes also a three-dimensional microtubule network within the ivory matrix [16, 17]. The structure of ivory is therefore similar to the dentin of teeth and bones, including human ones [18-22].

The nucleation and growth of apatite-collagen/gelatin composite materials and the diagenesis of ivory have been studied in detail [23-28]. In these prior approaches to hydroxylapatite-gelatin-composite synthesis, the apatite is formed in situ as the calcium solutions reacted with phosphoric acid [29, 30]. Here we used directly powders of gelatin and hydroxylapatite as starting materials by a solution-based process for the synthesis of our synthetic ivory composite.

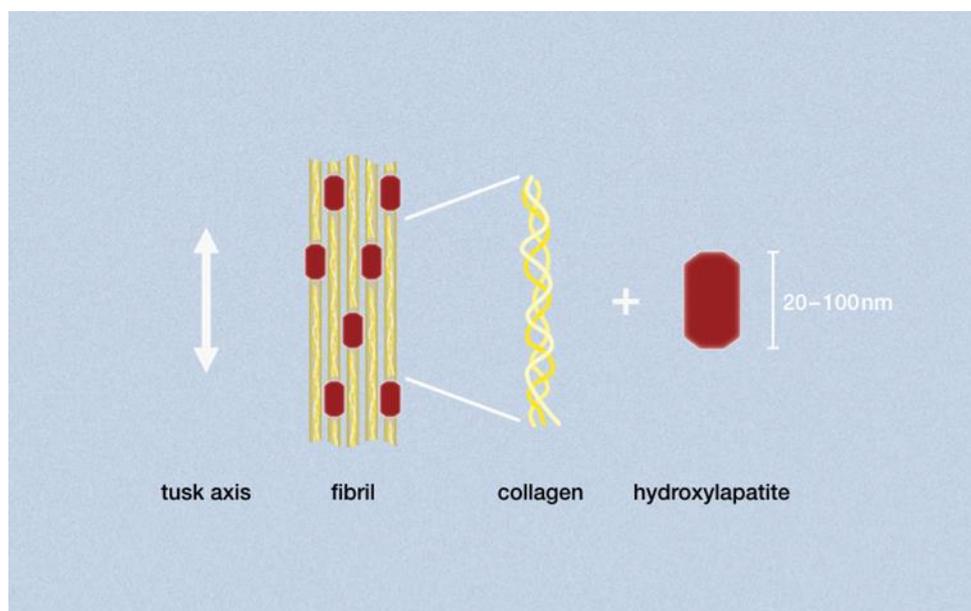

**Figure 1.** Sketch of the microstructure of natural ivory. This composite, dentin, is composed of a network of collagen fibrils oriented preferentially along the axis of the tusk, and rod-shaped hydroxylapatite crystals with a typical length of ~20–100 nm (figure adapted from [13] and [17]).

## 2. Materials and Methods

### 2.1. *Preparation*

Synthetic ivory: Samples were fabricated in a lab-scale beaker process with a gelatin-hydroxylapatite ratio (Gel/HA) of 1/3. Thus, 10 g of gelatin (type A from pork skin, granules, bloom 250–300 g, Biomol GmbH, Hamburg, Germany) was dissolved in 230 g of demineralized water, and 30 g of hydroxylapatite ($Ca_5[PO_4]_3OH$ powder, PA, Acros organics BVBA, Geel, Belgium) was suspended in 230 g of ethanol. Both were heated to 60°C while stirring, poured together and reduced to around 300 ml liquid (5 h), poured into flat boxes, and subsequently dried at ambient conditions for 5 days. The samples were kept under light pressure while drying to prevent bending. Afterwards, the samples were mechanically machined into plates. All samples presented in this paper were fabricated with pure hydroxylapatite. We find no improvement of functional properties achieved by embedding Mg-ions (e.g.).

Aging of synthetic ivory by glyoxal exposure: One plate of synthetic ivory (25×20×2 mm$^3$) was placed in a 1% glyoxal solution for 18 h (Gly-I) and another plate in a 4% glyoxal solution for 24 h (Gly-II). The samples were then removed from the solution and dried.

Natural ivory: Samples were from a plate (90×40×2 mm$^3$) of an African elephant tusk (Sudan); J. Schott Elfenbeinschnitzerei u. Mammutwerkstatt, Erbach, Germany.

Ivory substitutes for piano keys: 1: Ivoplast-white samples (plastic with filler) were from a plate (70×70×2 mm$^3$) used for key surfaces at Carl Sauter Pianofortemanufaktur GmbH & Co. KG, Spaichingen, Germany. 2: Elforyn samples (mineralized resin) were from a plate (160×25×2 mm$^3$) provided by Bachmann Kunststoff Technologien GmbH, Rödermark, Germany.

## 2.2. Characterization

Infrared spectra: IR spectra were recorded on an attenuated total-reflectance infrared spectrometer (ATR-IR, Spectrum Two, Perkin Elmer, Germany) equipped with a diamond window. Flat samples were pressed onto the window during the measurements.

Raman measurements: Raman spectra were measured on a laser-microscope Raman spectrometer (iHR 550 spectrometer; BXFM microscope manufactured by HORIBA, Bensheim, Germany) with confocal geometry. The three-grating spectrometer is equipped with an internal filter wheel and a Peltier-cooled CCD camera (Synapse). The resolution of the spectrometer (grating 1800 L/mm) is 1 wavenumber (cm$^{-1}$). Spectra were taken in quasi-backscattering geometry using the linearly polarized 532-nm line with power less than 100 mW, focused by a 100× objective onto the top surface of the sample.

X-ray powder diffraction: XRD patterns were measured in reflection mode with an Empyrean diffraction system (Malvern Panalytical BV, detector pixcel 3D, EA Almelo, Netherlands) CuK$\alpha$ radiation. The lattice constants were refined by the Rietveld method using crystal structure data of hydroxylapatite [31] via the TOPAS software (TOPAS Vers. 4.2, Brucker AXS) [32]. The crystallite sizes were calculated from the profile shape analysis (TOPAS) with the Scherrer equation.

Solid-state nuclear magnetic resonance measurements: $^1$H, $^{13}$C and $^{31}$P solid-state NMR spectra were collected on a Bruker Avance III 400 MHz instrument (magnetic field of 9.4 T) at the Larmor frequencies of 400.13, 100.61 and 161.98 MHz, respectively (Bruker GmbH, Rheinstetten, Germany). All spectra were obtained under conditions of Magic Angle Spinning (MAS) using a Bruker BL4 double- resonance MAS probe and 4-mm OD ZrO$_2$ spinners. $^1$H spectra were acquired in a simple Bloch decay experiment (i.e., pulse acquisition), with a π/2 pulse of 4.5 μs and an acquisition delay of 5 s, sufficient for a complete spin-lattice relaxation. $^{13}$C spectra were acquired using ramped cross polarization (CP) accompanied by composite proton decoupling with the proton RF power of 50 kHz [33]. $^{31}$P spectra were obtained from a Bloch decay experiment with composite proton decoupling and a relaxation delay of 300 s, which proved to be adequate for quantitative relaxation. Chemical shifts were referenced externally relative to TMS ($^1$H and $^{13}$C, δiso = 0.0 ppm), and 85% H$_3$PO$_4$ ($^{31}$P, δiso = 0.0 ppm).

Scanning electron microscopy (SEM): SEM images were acquired with a field-emission microscope (Merlin, Zeiss GmbH, Oberkochen, Germany) using an accelerating voltage of 1.50 kV and an Everhart–Thornley detector (SE2).

Transmission electron microscopy: High-resolution transmission electron microscope (HR-TEM) images were performed at 200 kV by a JEOL-ARM 200F TEM (JEOL, Tokyo, Japan) equipped with a cold field-emission gun and a CETCOR image corrector (CEOS Co. Ltd.). For ultramicrotomy specimen preparation, see [20]. The trimmed and cut samples were placed on Cu grids covered with a lacey carbon film (300 mesh).

Thermal conductivity: We used the thermal conductivity option of a Physical Properties Measurement System (PPMS, Quantum Design, Inc., San Diego, USA) to measure a slap-lined sample with dimensions of 6×6×3 mm$^3$. Contacts to copper leads were made using silver epoxy (Epo-Tek H20E), then cured in air at 80°C for 3 hours. Thermal conductivity was measured using the one heater–two thermometers technique with the steady-state pulsed method.

Hardness measurements: Hardness measurements were performed with a Nanoindenter XP system (KLA Tencor, USA) in a continuous stiffness mode (CSM) with a Berkowich tip. The maximum penetration depth was 3 μm. Young's modulus and hardness were calculated according to the Oliver and Pharr method [34] using data obtained for the range of the penetration depth between 2 and 3 μm.

Water absorption tests: Water absorption tests were performed on six materials: natural and synthetic (raw materials), synthetic aged by glyoxal (Gly-I, Gly-II), plastic Ivoplast, and mineralized resin Elforyn. The plate samples had dimensions of approximately 25×20 $mm^2$ and were 2–3 mm thick. The samples were weighed, baked in an oven at 100°C for 24 hours, then weighed again. Once the materials had cooled, the samples were placed in demineralized water. The samples were removed from the water and dried, then their mass was measured after 0.5, 1, 2, 5, and 24 hours. The masses are then normalized by the initial weight of the material after baking.

Grip tests: To simulate a finger pressing on a piano key, an artificial fingertip was constructed to measure the kinetic friction between it and the piano key. The fingertip construction was a flat, round rubber tip of polyurethane with a hardness of A 60–70, diameter of 12 mm and thickness of 3.5 mm (Article number 356.27.413, Häfele, Nagold, Germany) covered with leather. The leather was also cut to a diameter of 12 mm and then glued to the polyurethane tip. We used five different leathers to simulate different finger grips. All leathers were cowhide or lamb hide treated by a mineral tanning process (Table 4). The measurement setup consisted of the fingertip attached to a steel rod with a final mass of 0.5 kg, which was then positioned vertically on the material. The fingertip is constrained to move only vertically. The different test materials were glued to wooden blocks, each of which was placed on horizontal steel rods with rotational degrees of freedom. On one side of each wooden block, a force probe (FB-50, PCE Instruments, Meschede, Germany) with an x-axis motorized stage (MT-1/M-Z8, Thorlabs, Dachau, Germany) applied a horizontal force to the block. A constant velocity was applied at 0.1 mm/s for a distance of 2 mm. Force vs. time was recorded by the software provided by the force probe, which yielded an average value of the flat (kinetic friction) section of the force vs. time graph. The coefficient of the kinetic friction, $\mu_k$, is plotted on the radar plot for each material type and fingertip material.

**Table 4.** Leather types used for the grip measurements.

| Leather | Surface | Leather type | Origin | Thickness (mm) |
| --- | --- | --- | --- | --- |
| 1 | smooth | Nappa | Cow | 1.0–1.2 |
| 2 | rough | Nappa | Cow | 1.2–1.4 |
| 3 | smooth | Nappa | Lamb | 0.6–0.7 |
| 4 | rough | Nubuck | Cow | 1.0–1.2 |
| 5 | smooth | Nubuck | Cow | 1.0–1.2 |

## 3. Results and Discussion

### 3.1. *Development and Preparation*

To produce a material that is chemically identical to natural ivory, we used a simple solution-based process to synthesize the material directly, starting with hydroxylapatite ($Ca_5(PO_4)_3OH$) and gelatin as illustrated in Figure 2. Both, high-purity hydroxylapatite and gelatin powders are commercially available products. Gelatin is obtained by thermally hydrolyzing collagen from porcine skins and bones. With a chemical makeup that is closely related to that of collagen, gelatin is water-soluble and chemically suitable for the synthesis process, at the end of which the gelatin can be cross-linked to form collagen-like structures similar to those found in ivory. In this process, the gelatin expands to provide a matrix that is stabilized by hydroxylapatite, as described in Section 2.1. The properties of the resultant materials can be tuned by adjusting the hydroxylapatite/gelatin ratio, the gel strength of the gelatin, synthesis temperature, and solvent mixtures. The obtained slurry is dried to yield a bulk mass (see Figure 3a) or is applied as a coating, for example on wood.

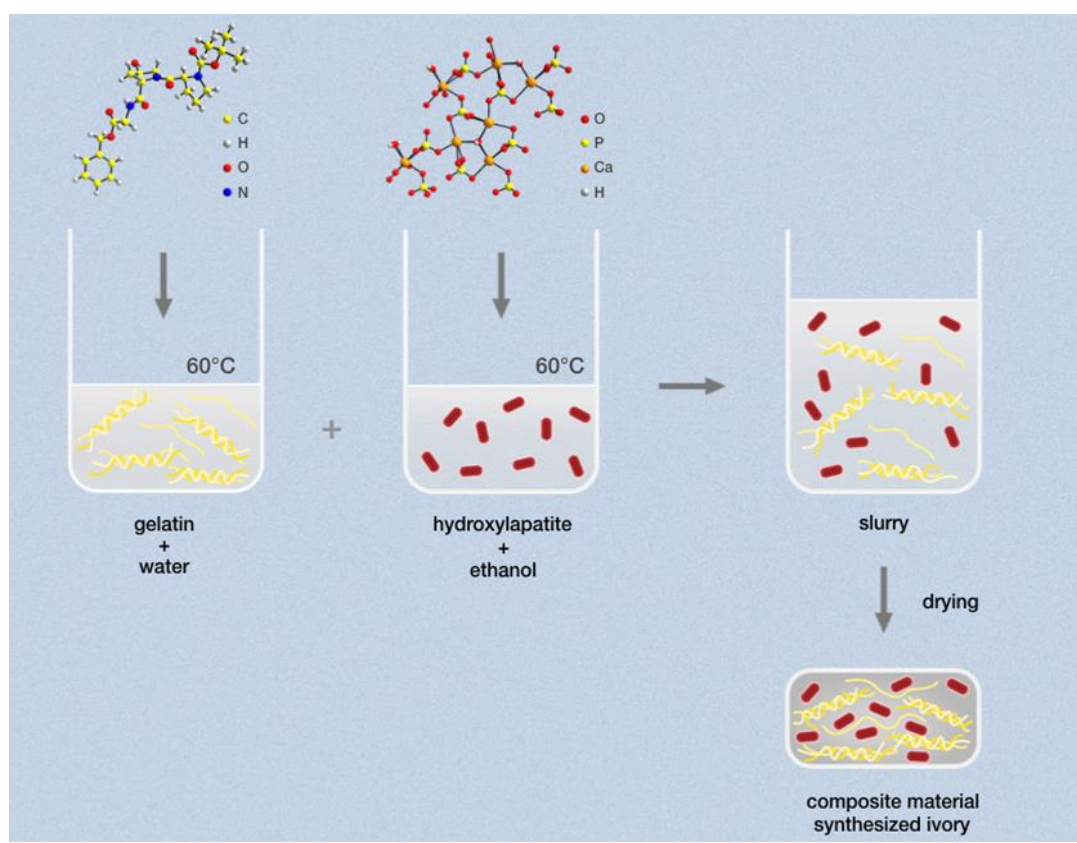

**Figure 2.** Schematic illustration of the synthesis of the gelatin-hydroxylapatite composite material. The synthesized ivory is formed from a slurry obtained by mixing the precursor solutions, followed by a drying process. The atomic structure of the two components are presented by segments of the collagen amino acid sequence (adapted from[35] and of the hydroxylapatite crystal structure (adapted from[31]).

During the fabrication process, pigments can be incorporated into the matrix to achieve virtually any color, including color patterns (see Figure 3b). In the same manner, additives such as antibacterial compounds or markers yielding fluorescent or DNA-based fingerprints for identification of the material may be added. Moreover, the surface of the material may be polished and exposed to cross-linking agents such as alum or glyoxal solutions, which allows the material's hardness to be optimized, makes it easy to clean, and determines its swelling capabilities (see Figure 3c).

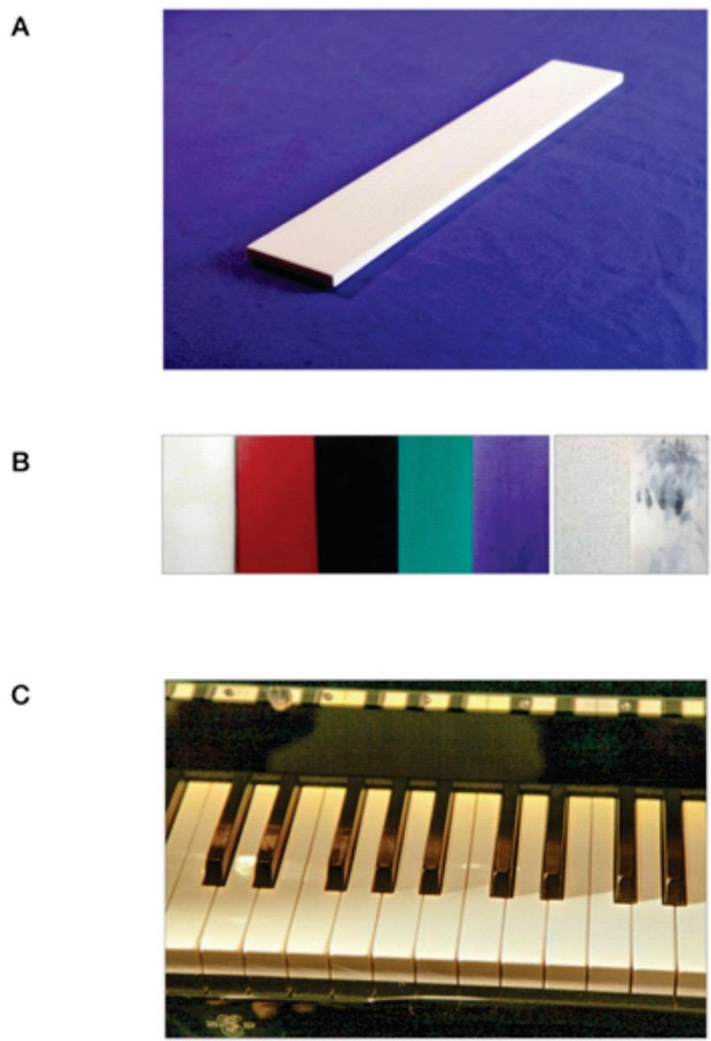

**Figure 3.** Samples of the composite material. a) machined plate (155×25×3 mm³) of pure synthesized ivory as fabricated; b) samples with colors and textures tuned by additives added into the slurry; c) piano keys with veneers consisting of synthesized ivory (a,c: photos by H.-J. Mannhart).

3.2. *Properties of Synthesized vs. Natural Ivory*

The synthesized product has a chemical composition within the spectrum of the compositions of natural ivories. The material is sturdy, hard, and ivory in color, and it readily absorbs moisture. Like natural ivory, the composite is machinable both by hand and machine in dry processes; it can be cut, sawed, milled, drilled, lathed, ground, and polished (Figure 3). Much like natural ivory, the synthesized material features a hardness of 0.4 GPa, a Young's modulus of 13 GPa (measured by nano-indention), and a density of 1.6 g/cm³. These values correspond very well to those of natural ivory (see Table 1). Only its thermal conductivity of ~0.8 W/Km at 25°C differs significantly: It is twice that of natural ivory, which is preferable for some uses, such as for piano keys. By design, synthesized ivory is isotropic, whereas natural ivory features a characteristic growth direction. This mechanical property is selected because it lends itself to machining.

The microstructure of the synthesized composite is characterized by the amorphous gelatin matrix with embedded hydroxylapatite rods typically measuring 10 nm × 50 nm as shown by the electron microscope images in Figure 4. The structural arrangements of these two composites differ qualitatively, as the synthesized material is isotropic and lacks the pores associated with the growth of natural ivory. We did not strive to reproduce the microstructure of natural ivory because it is not a parameter required to yield the functional properties relevant for our use case of piano keys, as we describe below.

**Table 1.** Comparison of mechanical and thermal properties of natural and synthesized ivory (standard deviation in parentheses).

| Sample | Hardness (GPa) | Young's modulus (GPa) | Thermal conductivity (W/K) at RT | Density (g/cm$^3$) |
|---|---|---|---|---|
| Synthesized ivory | 0.4 ± 0.1 | 13 ± 2 | 0.83 ± 0.02 | 1.6 ± 0.1 |
| Natural ivory | 0.36 [36] | 12.5 [37] | 0.34–0.5 [38] | 1.70 [37, 38] |

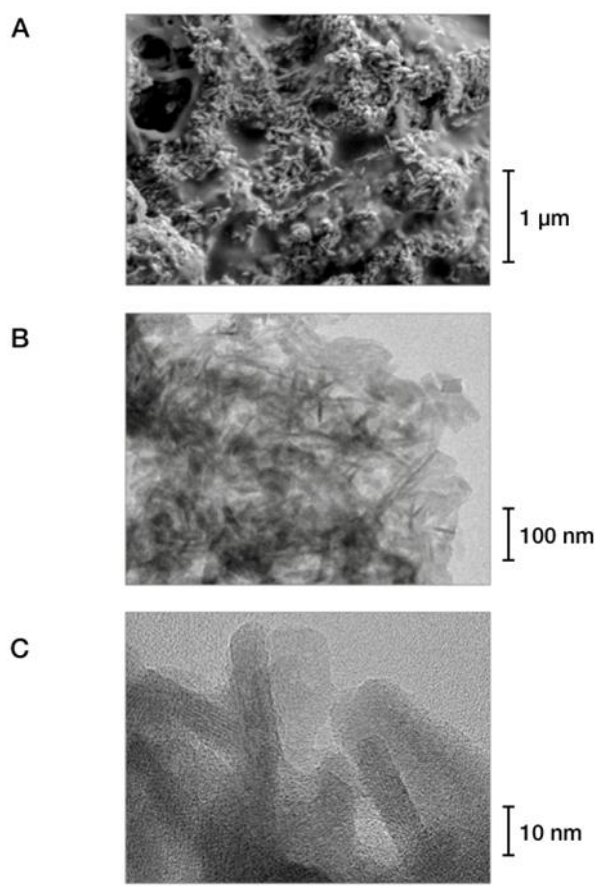

**Figure 4.** Microstructure of typical samples of synthesized ivory consisting of a gelatin-hydroxylapatite matrix. a) scanning electron microscopy image; b), c) transmission electron microscopy images; b, c) showing on two length scales the hydroxylapatite rods embedded in the gelatin matrix.

The x-ray diffraction patterns of the natural and synthesized composites are very similar, as shown by the x-ray powder diagrams in Figure 5a. The patterns of both materials are equivalent. The only significant difference are the pronounced series of diffraction peaks in the synthesized material caused by a larger crystallite size of the hydroxylapatite, which was predetermined by the crystallinity of the apatite powder used as a reagent. The Rietveld refinements of the powder patterns reveal the same lattice constants for both materials, which also correspond very well to the lattice constants of the hydroxylapatite powder; cf. Table 2 for refined lattice constants and crystallite sizes.

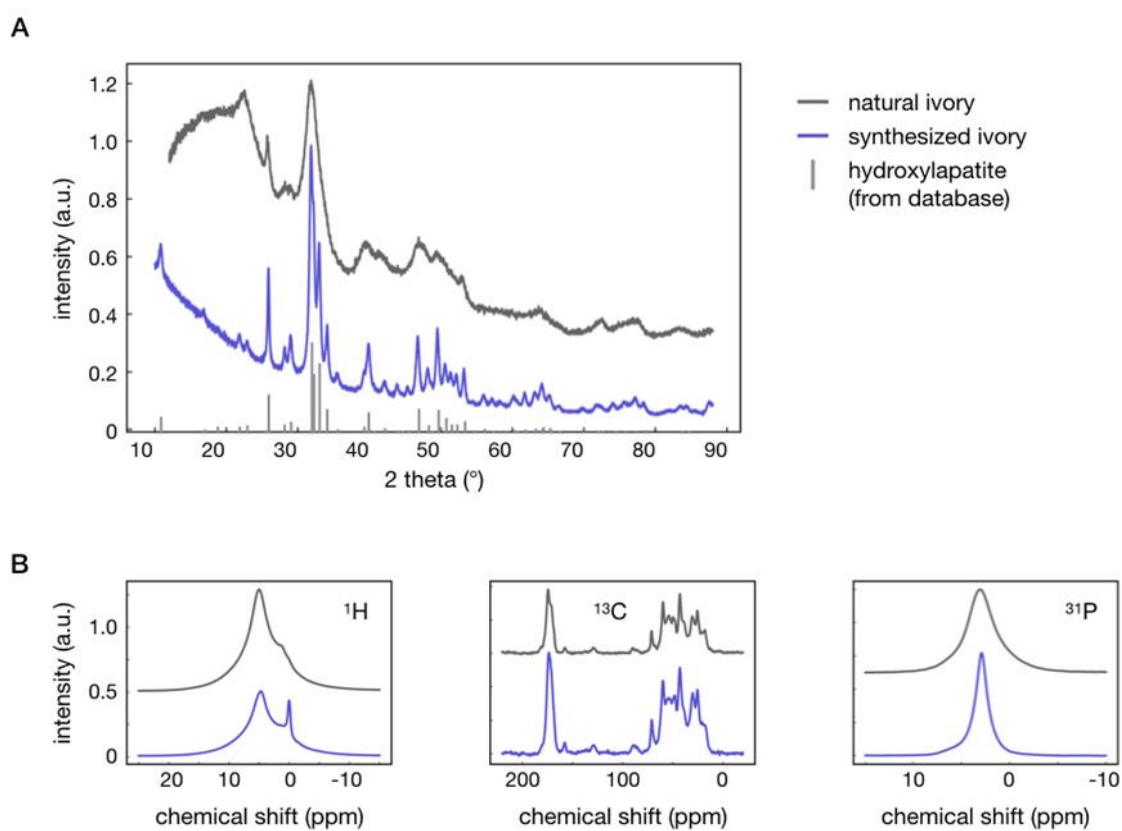

**Figure 5.** a) X-ray powder diffraction patterns and b) solid-state NMR spectra of natural and synthesized ivory.

**Table** 2 Lattice parameters (a, c) and crystallite size (λ) of natural and synthesized ivory compared to hydroxylapatite powder [32] (standard deviation in parentheses).

| Sample | a (Å) | c (Å) | λ (Å) |
| --- | --- | --- | --- |
| Synthesized ivory | 9.4202 ± 0.0004 | 6.8897 ± 0.0003 | 220 ± 1 |
| Natural ivory | 9.415 ± 0.006 | 6.876 ± 0.005 | 60 ± 1 |
| Hydroxylapatite | 9.4222 ± 0.0001 | 6.8812 ± 0.0001 | 1030 |

Solid-state nuclear magnetic resonance (NMR) provides additional information on the synthesized composite. Figure 5b compares the $^1$H, $^{13}$C, and $^{31}$P solid-state NMR spectra of the natural and synthesized materials (for more details, see Section 2.2). The spectra are virtually identical. The single exception is that the $^1$H signal at ~0 ppm is visible for the synthesized ivory but absent for natural ivory, which is caused by structural OH-groups of the hydroxylapatite. Whereas the peak is suppressed by carbonated apatite in this specific sample of natural ivory, it is exhibited clearly in numerous other ivory specimens [39]. The smaller peak width in the $^{31}$P spectrum is caused by the higher structural order of the hydroxylapatite in the synthesized material. The two materials are completely indistinguishable by Raman scattering spectroscopy (Figure 6a), and have almost identical IR spectra (Figure 6b). The higher water content of the natural ivory is expressed by the broad band between 3000 and 3500 cm$^{-1}$ (arrow), which can arise from carbonated apatite in natural ivory. Carbonate groups in the apatite of natural ivory, which we did not strive to mimic, result in distinct bands at 871 and 1413/1446 cm$^{-1}$ for natural ivory. The vibration frequencies of the IR and Raman bands of both compounds are listed in Table 3.

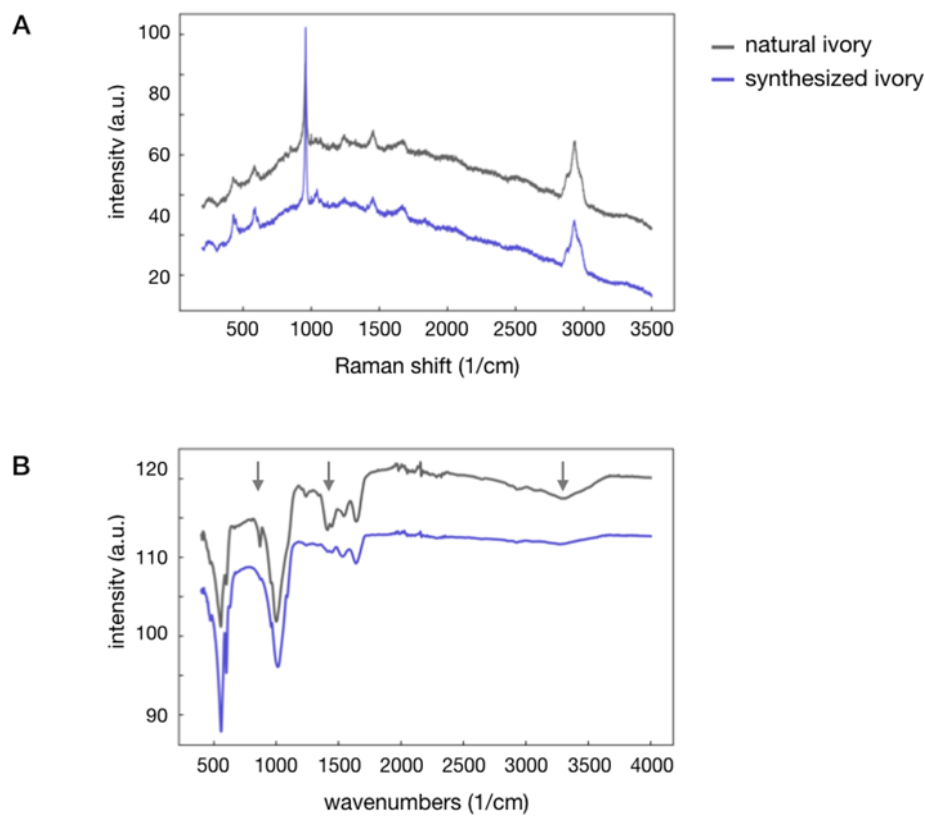

**Figure 6.** a) Raman and b) infrared spectra of natural and synthesized ivory. The arrows in b) highlight spectral features that result from the rather high carbonate content (left and middle arrow) and the high water content (right arrow) in this sample of natural ivory.

In summary, the results of our analytical investigations reveal that natural and synthesized ivory are virtually identical, except for those parameters corresponding to impurities or microstructures of specific samples of natural ivory that are not relevant for the envisioned use case—and that we therefore did not aim to reproduce—or that differ by design in order to produce superior physical properties for that use case.

**Table 3.** Relevant infrared frequencies and Raman shifts including vibrational assignments of natural and synthesized ivory.

| | IR frequency | (cm$^{-1}$) | | Raman shift | (cm$^{-1}$) |
|---|---|---|---|---|---|
| Synthesized ivory | Natural ivory | Assignments | Synthesized ivory | Natural ivory | Assignments |
| 471 | 470 | δ PO$_4$ | 428/448 | 429 | δ PO$_4$ |
| 557 | 555 | δ PO$_4$ | 578/589 | 584 | δ PO$_4$ |
| 599 | 600 | δ PO$_4$ | | | |
| 876 | 871 | ν CO$_3$ | | | |
| 961 | 960 | ν PO$_4$ | 959 | 960 | ν PO$_4$ |
| 1014 | 1002 | ν PO$_4$ | 1001 | 1000 | ν CC (ar) |
| 1239 | 1240 | δ NH | 1043 | 1043/1067 | ν PO$_4$, ν CO$_3$ |
| 1413 | 1413 | ν CO$_3$ | 1245 | 1246 | δ NH |
| 1445 | 1446 | ν CO$_3$ | 1452 | 1457 | δ NH, δ CH$_2$ |
| 1534 | 1544 | δ NH | 1672 | 1676 | ν C=O, δ NH |
| 1640 | 1645 | δ NH | | | |
| | ≈ 3300 | ν OH (H$_2$O) | 2877/2933 | 2880/2937 | ν CH$_2$, ν CH$_3$ |

ν: stretching, δ: bending, ar: aromatic.

3.3. *Requirements for Replacing Natural-Ivory Piano Keys*

For piano key surfaces, the relevant functional properties are moisture absorption, friction to fingers (or "grip"), color, ease of cleaning, robustness, and long-term stability. No existing ivory substitute can compete in all these aspects. Indeed, most of them are inferior with respect to moisture absorption and grip. Figure 7 and 8 compare moisture absorption and grip, respectively, of the synthesized ivory composite with natural ivory and the currently used substitute materials Ivoplast-white (plastic with filler) and Elforyn (mineralized resin). In Figure 7, the amount of water absorbed by these materials immersed in demineralized water at 300 K is shown as a saturation curve, revealing a relative mass increase of natural ivory of 28% after 24 hours. The water uptake of Elforyn and Ivoplast reaches only 3% and 0.5%, respectively. Our synthesized ivory composite achieves a relative water absorption of 72%, a factor of 2.5 higher than natural ivory. Moreover, the water absorption capability of synthesized ivory is tunable by crosslinking the gelatin chains with glyoxal. These crosslinks are formed by reactions of the amine functions of the amino acids with the aldehyde function of the glyoxal reagent [40]. By adjusting the crosslinker solution concentration and the exposure time, the water absorption of the synthesized ivory can be tuned to match the value of natural ivory (Figure 7, curves Gly-I and Gly-II).

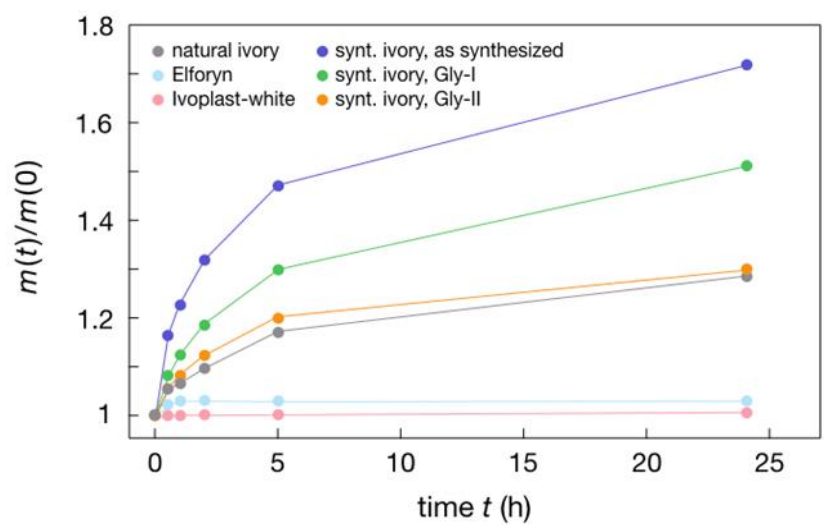

**Figure 7.** Comparison of the water absorption of natural ivory and of synthesized ivory with different surface treatments. The figure shows the water uptake of samples immersed in water at room temperature as function of time. For measurement details, see Materials and Methods.

To analyze the grip properties, we measured the friction forces between composite samples and artificial fingers made of various fingertip materials (for details, see Section 2.2). Figure 8 shows that tuning the surface properties of the synthesized ivory using crosslinker solutions yields a broad range of friction coefficients, including those of natural ivory, in agreement with the subjective observations of professional pianists. By tuning the friction coefficients of the surface during the fabrication process, it would therefore be possible to match the grip of keys to the preferences of individual pianists.

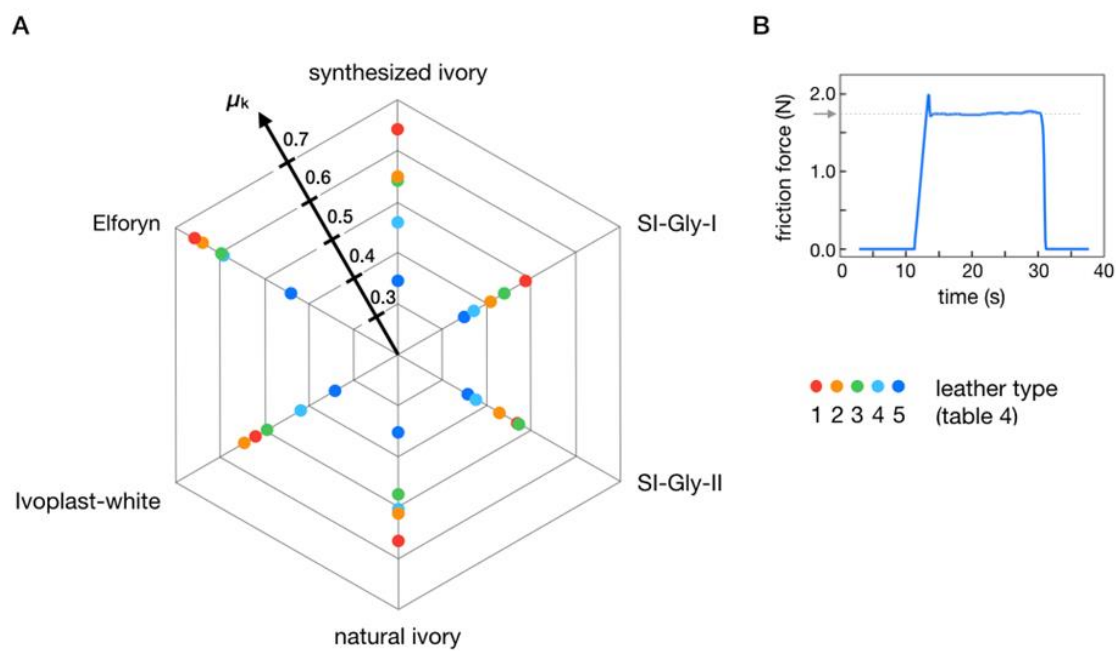

**Figure 8.** a) Comparison of the lateral grip forces (kinetic friction $\mu_k$) acting between artificial test fingers and samples of natural and synthesized ivory tuned by surface treatments as specified in the Materials and Methods Section. The test fingers are covered with several kinds of leather (cf. Table 4) and are pressed against the samples over a contact area of several mm² with an approximate 5 N force. The coefficient of the kinetic friction $\mu_k$ is obtained from the plateau values of the friction force vs. time curve of b). b) shows the friction force vs. time curve for SI-Gly-II sample with leather 5. For measurement details, see Section 2.2.

The addition of pigments makes it possible to achieve synthesized ivory to match the color gamut of natural ivory. Moreover, the surfaces of natural and synthetic ivory are optically so similar that it is virtually impossible to perceive a difference. The synthesized composite is as robust and easy to clean as natural ivory, which we tested over a 6-month period with a piano equipped with keys of synthetic ivory (Figure 3c).

## 4. Conclusions

For good reasons, ivory has traditionally been the material of choice for piano key surfaces. However, as it has become largely unavailable due to the vulnerable status [41] and corresponding protection of elephants, there is demand for an alternative, a sustainable material with comparable functional properties. To obtain such a material, we synthesized a composite that is chemically identical to natural ivory, nontoxic and non-hazardous to living tissue. We then optimized its functional properties, disregarding irrelevant microstructural and trace mineral differences from those of natural ivory. The raw materials used to synthesize the ivory substitute are abundantly available and economical. At the end of its lifetime, the composite decays into environmentally friendly apatite and gelatin waste products. Possible applications extend beyond our use case of piano keys and could conceivably include jewelry, and alternatives to hydrocarbon-based plastics for applications that require superior tactile properties or bio-degradability.

**Author Contributions:** D.F. conceived the experimental procedure, prepared the specimens, performed the experiments, analyzed and elaborated the results, provided discussion and wrote and edited portions of the paper. S.C.P. prepared the specimens, performed the experiments, analyzed and elaborated the results, provided discussion and wrote and edited portions of the paper. J.M. provided discussion and elaboration of results, wrote and edited portions of the paper.

**Funding:** This research received no external funding.

**Acknowledgments:** We gratefully acknowledge helpful discussions with and technical support by M.-L. Schreiber (IR measurements), H. Hoier (XRD), V. Duppel (SEM), V. Srot (HR-TEM), B. Bußmann (TEM sample preparation), I. Moudrakovski (NMR), J. Bruin, Y. Zhang (thermal conductivity), I. Hagel, M. Schmid, A. Ortlieb (machining of synthesized ivory), all Max Planck Institute of Solid State Research, Z. Burghard (hardness measurements) University Stuttgart, Institute of Material Science, and Carl Sauter Pianofortemanufaktur GmbH & Co. KG (piano keys manufacturing).

**Conflicts of Interest:** The further development and commercialization of synthetic ivory is performed in a startup company (ivortec, co-owners: Max Planck Society, JM).

References

1. Chase, M. J.; Schlossberg, S.; Griffin, C. R.; Bouche, P. J. C.; Djene, S. W.; Elkan, P. W.; Ferreira, S.; Grossman, F.; Kohi, E. M.; Landen, K.; Omondi, P.; Peltier, A.; Selier, S. A. J.; Sutcliffe, R. Continent-wide survey reveals massive decline in African savannah elephants. *Peerj* **2016**, *4*, 24.
2. Biggs, D.; Holden, M. H.; Braczkowski, A.; Cook, C. N.; Milner-Gulland, E. J.; Phelps, J.; Scholes, R. J.; Smith, R. J.; Underwood, F. M.; Adams, V. M.; Allan, J.; Brink, H.; Cooney, R.; Gao, Y. F.; Hutton, J.; Macdonald-Madden, E.; Maron, M.; Redford, K. H.; Sutherland, W. J.; Possingham, H. P. Breaking the deadlock on ivory. *Science* **2017**, *358*, 1378-1381.
3. Wittemyer, G.; Northrup, J. M.; Blanc, J.; Douglas-Hamilton, I.; Omondi, P.; Burnham, K. P. Illegal killing for ivory drives global decline in African elephants. *Proc. Natl. Acad. Sci. U. S. A.* **2014**, *111*, 13117-13121.
4. Espinoza EO, M. M. Identification guide for ivory and ivory substitutes. *World Wildlife Fund and The Conservation Foundation* 1991, pp 1-38.
5. Muller, K.; Reiche, I. Differentiation of archaeological ivory and bone materials by micro-PIXE/PIGE with emphasis on two Upper Palaeolithic key sites: Abri Pataud and Isturitz, France. *J. Archaeol. Sci.* **2011**, *38*, 3234-3243.
6. Chu, Y. H.; Meyers, M. A.; Wang, B.; Yang, W.; Jung, J. Y.; Coimbra, C. F. M. A Sustainable Substitute for Ivory: the Jarina Seed from the Amazon. *Sci. Rep.* **2015**, *5*, 10.
7. Vollrath, F.; Mi, R. X.; Shah, D. U. Ivory as an Important Model Bio-composite. *Curator-the Museum Journal* **2018**, *61*, 95-110.
8. Braun, D. Polymer research before Hermann Staudinger The Long Road to Macromolecule. *Chem. Unserer Zeit* **2012**, *46*, 310-320.
9. Feldman, D. Polymer history. *Des. Monomers Polym.* **2008**, *11*, 1-15.
10. Germershaus, O.; Luhmann, T.; Rybak, J. C.; Ritzer, J.; Meinel, L. Application of natural and semi-synthetic polymers for the delivery of sensitive drugs. *Int. Mater. Rev.* **2015**, *60*, 101-130.
11. Eder, M.; Amini, S.; Fratzl, P. Biological composites-complex structures for functional diversity. *Science* **2018**, *362*, 543-547.
12. Su, X. W.; Cui, F. Z. Hierarchical structure of ivory: from nanometer to centimeter. *Materials Science & Engineering, C: Mater. Bio. Appl.* **1999**, *7*, 19-29.
13. Jantou-Morris, V.; Horton, M. A.; McComb, D. W. The nano-morphological relationships between apatite crystals and collagen fibrils in ivory dentine. *Biomaterials* **2010**, *31*, 5275-5286.
14. Singh, R. R.; Goyal, S. P.; Khanna, P. P.; Mukherjee, P. K.; Sukumar, R. Using morphometric and analytical techniques to characterize elephant ivory. *Forensic Sci. Int.* **2006**, *162*, 144-151.
15. Buddhachat, K.; Thitaram, C.; Brown, J. L.; Klinhom, S.; Bansiddhi, P.; Penchart, K.; Ouitavon, K.; Sriaksorn, K.; Pa-in, C.; Kanchanasaka, B.; Somgird, C.; Nganvongpanit, K. Use of handheld X-ray fluorescence as a non-invasive method to distinguish between Asian and African elephant tusks. *Sci. Rep.* **2016**, *6*, 11.
16. Alberic, M.; Dean, M. N.; Gourrier, A.; Wagermaier, W.; Dunlop, J. W. C.; Staude, A.; Fratzl, P.; Reiche, I. Relation between the Macroscopic Pattern of Elephant Ivory and Its Three-Dimensional Micro-Tubular Network. *Plos One* **2017**, p 22.

17. Alberic, M.; Gourrier, A.; Wagermaier, W.; Fratzl, P.; Reiche, I. The three-dimensional arrangement of the mineralized collagen fibers in elephant ivory and its relation to mechanical and optical properties. *Acta Biomater.* **2018,** *72*, 342-351.

18. Seidel, R.; Gourrier, A.; Kerschnitzki, M.; Burghammer, M.; Fratzl, P.; Gupta, H. S.; Wagermaier, W. Synchrotron 3D SAXS analysis of bone nanostructure. In *Bioinspired Biomimetic and Nanobiomaterials*, Ice Publishing: Westminister, **2012**; Vol. 1, pp 123-131.

19. Busch, S.; Schwarz, U.; Kniep, R. Morphogenesis and structure of human teeth in relation to biomimetically grown fluorapatite-gelatine composites. *Chem. Mater.* **2001,** *13*, 3260-3271.

20. Srot, V.; Bussmann, B.; Salzberger, U.; Koch, C. T.; van Aken, P. A. Linking Microstructure and Nanochemistry in Human Dental Tissues. *Microsc. Microanal.* **2012,** *18*, 509-523.

21. Olszta, M. J.; Cheng, X. G.; Jee, S. S.; Kumar, R.; Kim, Y. Y.; Kaufman, M. J.; Douglas, E. P.; Gower, L. B. Bone structure and formation: A new perspective. *Mater. Sci. Eng. R: Rep.* **2007,** *58*, 77-116.

22. Dunlop, J. W. C.; Fratzl, P. Biological Composites. In *Annual Review of Materials Research, Vol 40*, Clarke, D. R., Ruhle, M.; Zok, F., Eds. Annual Reviews: Palo Alto, **2010**; Vol. 40, pp 1-24.

23. Tlatlik, H.; Simon, P.; Kawska, A.; Zahn, D.; Kniep, R. Biomimetic fluorapatite-gelatine nanocomposites: Pre-structuring of gelatine matrices by ion impregnation and its effect on form development. *Angew. Chem., Int. Ed. Engl.* **2006,** *45*, 1905-1910.

24. Simon, P.; Rosseeva, E.; Buder, J.; Carrillo-Cabrera, W.; Kniep, R. Embryonic States of Fluorapatite-Gelatine Nanocomposites and Their Intrinsic Electric-Field-Driven Morphogenesis: The Missng Link on the Way from Atomistic Simulations to Pattern Formation on the Mesoscale. *Adv. Funct. Mater.* **2009,** *19*, 3596-3603.

25. Wang, Y.; Azais, T.; Robin, M.; Vallee, A.; Catania, C.; Legriel, P.; Pehau-Arnaudet, G.; Babonneau, F.; Giraud-Guille, M. M.; Nassif, N. The predominant role of collagen in the nucleation, growth, structure and orientation of bone apatite. *Nat. Mater.* **2012,** *11*, 724-733.

26. Bleek, K.; Taubert, A. New developments in polymer-controlled, bio-inspired calcium phosphate mineralization from aqueous solution *Acta Biomater.* **2013,** *9*, 8466-8466.

27. Elsharkawy, S.; Al-Jawad, M.; Pantano, M. F.; Tejeda-Montes, E.; Mehta, K.; Jamal, H.; Agarwal, S.; Shuturminska, K.; Rice, A.; Tarakina, N. V.; Wilson, R. M.; Bushby, A. J.; Alonso, M.; Rodriguez-Cabello, J. C.; Barbieft, E.; Hernandez, A. D.; Stevens, M. M.; Pugno, N. M.; Anderson, P.; Mata, A. Protein disorder-order interplay to guide the growth of hierarchical mineralized structures. *Nat. Commun.* **2018,** *9*, 2145.

28. Alberic, M.; Gourrier, A.; Muller, K.; Zizak, I.; Wagermaier, W.; Fratzl, P.; Reiche, I. Early diagenesis of elephant tusk in marine environment. *Palaeogeogr. Palaeoclimatol. Palaeoecol.* **2014,** *416*, 120-132.

29. Kollmann, T.; Simon, P.; Carrillo-Cabrera, W.; Braunbarth, C.; Poth, T.; Rosseeva, E. V.; Kniep, R. Calcium Phosphate-Gelatin Nanocomposites: Bulk Preparation (Shape- and Phase-Control), Characterization, and Application as Dentine Repair Material. *Chem. Mater.* **2010,** *22*, 5137-5153.

30. Chang, M. C.; Douglas, W. H.; Tanaka, J. Organic-inorganic interaction and the growth mechanism of hydroxyapatite crystals in gelatin matrices between 37 and 80 degrees C. *J. Mater. Sci.: Mater. Med.* **2006,** *17*, 387-396.

31. Kay, M. I.; Young, R. A.; Posner, A. S. CRYSTAL STRUCTURE OF HYDROXYAPATITE. *Nature* **1964,** *204*, 1050-1052.

32. Coelho, A. A. *General Profile and Structure Analysis software for powder diffraction data*, Bruker AXS GmbH: Karlsruhe, Germany, **2009**.

33. MacKenzie, K. J. D.; Smith, M. E. *Multinuclear Solid-State NMR of Inorganic Materials*. **2002**.
34. Oliver, W. C.; Pharr, G. M. Measurement of hardness and elastic modulus by instrumented indentation: Advances in understanding and refinements to methodology. *J. Mater. Res.* **2004,** *19*, 3-20.
35. Doi, M.; Imori, K.; Sakaguchi, N.; Asano, A. Boc-Pro-Hyp-Gly-OBzl and Boc-Ala-Hyp-Gly-OBzl, two repeating triplets found in collagen. *Acta Cryst. Sect. C: Cryst. Struct. Commun.* **2006,** *62*, O577-O580.
36. Cui, F. Z.; Wen, H. B.; Zhang, H. B.; Li, H. D.; Liu, D. C. Anisotropic indentation morphology and hardness of Natural Ivory. *Mater. Sci.Eng. C* **1994,** *2*, 87-91.
37. Rajaram, A. Tensile properties and fracture of ivory. *J. Mater. Sci. Lett.* **1986,** *5*, 1077-1080.
38. Jakubinek, M. B.; Samarasekera, C. J.; White, M. A. Elephant ivory: A low thermal conductivity, high strength nanocomposite. *J. Mater. Res.* **2006,** *21*, 287-292.
39. Bracco, S.; Brajkovic, A.; Comotti, A.; Rolandi, V. Characterization of Elephant and Mammoth Ivory by Solid State NMR. *Periodico Di Mineralogia* **2013,** *82*, 239-250.
40. Duconseille, A.; Astruc, T.; Quintana, N.; Meersman, F.; Sante-Lhoutellier, V. Gelatin structure and composition linked to hard capsule dissolution: A review. *Food Hydrocolloids* **2015,** *43*, 360-376.
41. Thouless, C. R.; Dublin, H. T. B., J.J.;; Skinner, D. P.; Daniel, T. E.; Taylor, R. D.; Maisels, F.; Frederick, H. L.; Bouché, P. African Elephant Status Report 2016: an update from the African Elephant Database. *Occasional Paper Series of the IUCN Species Survival Commission* **2016**, p 309.